\documentclass[sigconf]{acmart}
\usepackage{mathtools}
\usepackage{subcaption}
\usepackage{float}
\usepackage{stfloats}
\usepackage{makecell}

\AtBeginDocument{%
  }

\copyrightyear{2025}
\acmYear{2025}
\setcopyright{acmlicensed}\acmConference[RecSys '25]{Proceedings of the Nineteenth ACM Conference on Recommender Systems}{September 22--26, 2025}{Prague, Czech Republic}
\acmBooktitle{Proceedings of the Nineteenth ACM Conference on Recommender Systems (RecSys '25), September 22--26, 2025, Prague, Czech Republic}
\acmDOI{10.1145/3705328.3748065}
\acmISBN{979-8-4007-1364-4/2025/09}

\begin{document}

\title{LONGER: Scaling Up Long Sequence Modeling in \\Industrial Recommenders}


\author{Zheng Chai*}
\thanks{*These authors contributed equally.}
\thanks{\textsuperscript{†}Corresponding Author.}
\email{chaizheng.cz@bytedance.com}
\affiliation{%
  \institution{ByteDance}
  \city{Hangzhou}
  \country{China}
}

\author{Qin Ren*}
\email{renqin.97@bytedance.com}
\affiliation{%
  \institution{ByteDance}
  \city{Beijing}
  \country{China}
}

\author{Xijun Xiao*}
\email{xiaoxijun@bytedance.com}
\affiliation{%
  \institution{ByteDance}
  \city{Beijing}
  \country{China}
}

\author{Huizhi Yang*}
\email{yanghuizhi@bytedance.com}
\affiliation{%
  \institution{ByteDance}
  \city{Beijing}
  \country{China}
}

\author{Bo Han*}
\email{hanbo.1990@bytedance.com}
\affiliation{%
  \institution{ByteDance}
  \city{Shanghai}
  \country{China}
}

\author{Sijun Zhang}
\email{zhangsijun.randyz@bytedance.com}
\affiliation{%
  \institution{ByteDance}
  \city{Beijing}
  \country{China}
}

\author{Di Chen}
\email{chendi.666@bytedance.com}
\affiliation{%
  \institution{ByteDance}
  \city{Beijing}
  \country{China}
}

\author{Hui Lu}
\email{luhui.xx@bytedance.com}
\affiliation{%
  \institution{ByteDance}
  \city{Hangzhou}
  \country{China}
}

\author{Wenlin Zhao}
\email{zhaowenlin@bytedance.com}
\affiliation{%
  \institution{ByteDance}
  \city{Beijing}
  \country{China}
}

\author{Lele Yu}
\email{yulele@bytedance.com}
\affiliation{%
  \institution{ByteDance}
  \city{San Jose}
  \country{USA}
}

\author{Xionghang Xie}
\email{xiexionghang@bytedance.com}
\affiliation{%
  \institution{ByteDance}
  \city{Beijing}
  \country{China}
}

\author{Shiru Ren}
\email{renshiru2000@gmail.com}
\affiliation{%
  \institution{ByteDance}
  \city{Beijing}
  \country{China}
}

\author{Xiang Sun}
\email{sunxiang.0@bytedance.com}
\affiliation{%
  \institution{ByteDance}
  \city{Beijing}
  \country{China}
}

\author{Yaocheng Tan}
\email{tanyaocheng@bytedance.com}
\affiliation{%
  \institution{ByteDance}
  \city{Beijing}
  \country{China}
}


\author{Peng Xu}
\email{xupeng@bytedance.com}
\affiliation{%
  \institution{ByteDance}
  \city{San Jose}
  \country{USA}
}

\author{Yuchao Zheng\textsuperscript{†}}
\email{zhengyuchao.yc@bytedance.com}
\affiliation{%
  \institution{ByteDance}
  \city{Beijing}
  \country{China}
}

\author{Di Wu}
\email{di.wu@bytedance.com}
\affiliation{%
  \institution{ByteDance}
  \city{Beijing}
  \country{China}
}

\renewcommand{\shortauthors}{Chai et al.}

\begin{abstract}
Modeling ultra-long user behavior sequences is critical for capturing both long- and short-term preferences in industrial recommender systems. Existing solutions typically rely on two-stage retrieval or indirect modeling paradigms, incurring upstream-downstream inconsistency and computational inefficiency. In this paper, we present \textbf{LONGER}, a Long-sequence Optimized traNsformer for GPU-Efficient Recommenders. LONGER incorporates (i) a global token mechanism for stabilizing attention over long contexts, (ii) a token merge module with lightweight InnerTransformers and hybrid attention strategy to reduce quadratic complexity, and (iii) a series of engineering optimizations, including training with mixed-precision and activation recomputation, KV cache serving, and the fully synchronous model training and serving framework for unified GPU-based dense and sparse parameter updates. LONGER consistently outperforms strong baselines in both offline metrics and online A/B testing in both advertising and e-commerce services at ByteDance, validating its consistent effectiveness and industrial-level scaling laws. Currently, LONGER has been validated and fully deployed across dozens of real-world influential scenarios at ByteDance, serving billions of users.
\end{abstract}





\maketitle

\section{Introduction}
In recommendation systems, ultra-long user historical behavior sequences comprehensively encapsulate both long-term and short-term user preferences\cite{he2023survey, de2021transformers4rec}. While early sequential modeling architectures have been extensively studied and widely adopted in both academia and industry, their applications remain largely confined to short-sequence scenarios (sequence lengths of $10^2 - 10^3$). Fully modeling long sequences (length > $10^3$) offers significant benefits for recommendation accuracy and diversity, and helps mitigate the information cocoon phenomenon. However, due to the computational constraints, current industry $de \ facto$ practices for long-sequence modeling primarily adopt the following strategies:

\begin{itemize}
    \item Two-stage retrieval. Select top-$k$ items (typically $k$ at $10^2$) from the original ultra-long sequence that are most relevant to the current candidate item, followed by end-to-end short sequence modeling. The most representative works include SIM\cite{pi2020search} and TWIN\cite{chang2023twin,si2024twin}.
    \item Pre-trained User Embeddings\cite{zhou2020s3,hou2022towards,liu2021augmenting}. In industry, it is a common practice to pre-train the entire ultra-long sequence in a source model and derive a condensed user embedding (UE), which can then be transferred to downstream recommendation models. Leveraging high-performance advanced GPUs, this method supports pre-training with sequence of up to $10^3$ length and multiple-layered transformers.
    \item Memory-augmented Models. The multi-channel user interest memory network (MIMN) \cite{pi2019practice} offers a neural Turing machine and memory induction unit-based structure for user sequence memorizing, and large memory network (LMN)\cite{lu2025large} presents a lightweight structure with product quantization-based decomposition. The memory augmented recommendation model (MARM) \cite{lv2024marm} proposes a memory-for-computation trade-off paradigm, which caches the intermediate results from computationally intensive modules.
\end{itemize}

\noindent While these strategies significantly improve computational efficiency, they inevitably sacrifice raw full-sequence information due to the upstream-downstream inconsistency or the indirect perception of the original ultra-long sequence, and thus these approaches essentially provide an intermediate stage in the evolution toward end-to-end long-sequence modeling. 


Recently, the rapid advancement of large language models, exemplified by GPT \cite{radford2018improving}, has established scaling laws - empirical principles predicting performance improvements with increased model size, data volume, and computility. These scaling laws have recently guided innovations in recommendation systems. For example, HSTU \cite{zhai2024actions} consists of a stack of identical self-attention layers connected by residual connections for modeling long sequences, which shows better performance than vanilla Transformer architectures. Wukong\cite{zhang2024wukong} develops a stacked factorization machine and linear compression block based architecture for interaction, and validates the scaling laws in recommendation.

At the same time, with the rapid advancements in computing infrastructure (e.g., GPU FLOPs/Memory, engineering large-scale computing platforms and frameworks), it has excitingly enabled us to pioneer an end-to-end ultra-long sequence modeling paradigm in industrial-scaled recommendation systems. Therefore, advancing end-to-end modeling of ultra-long sequences, along with continuously scaling sequence length and refining the architecture for long-sequence modeling, represents a critical imperative for next generation sequence modeling frameworks.

To this end, we propose the Long-sequence Optimized traNsformer for GPU-Efficient Recommenders, i.e., \textbf{LONGER}. In the framework, we organize the sequence input as the global tokens and raw sequences, based on which an inner-transformer based token merge methodology is developed for effectively reducing computing budget. Besides, as there is generally a lot of noise present in users' ultra long sequences, we utilize an efficient hybrid attention strategy for improving computational efficiency while maintaining model performance. Besides, to fully deploy LONGER at an industrial level with billion-user scale, we present a series of engineering optimizations, including a fully synchronous training and serving framework with mixed-precision and activation recomputation, and a KV cache serving strategy. Overall, the contributions are mainly summarized as follows:

\begin{itemize}
    \item We present LONGER, a long-sequence optimized transformer structure for GPU-efficient recommenders. It presents an industrial GPU-efficient viewpoint by optimizing transformer structures and scales up user sequence modeling length to 10,000 in an end-to-end manner in industry.
    \item LONGER sufficiently improves computational efficiency through token merge and hybrid attention strategies, which reduce \textasciitilde 50\% FLOPs and are validated to be almost lossless in performance. Besides, a fully-optimized industrial training and serving framework is devised for further improved GPU computational efficiency and online deployments.
    \item Thorough experiments are conducted to validate the efficacy. Offline experiments on a billion-scale industrial dataset, and online A/B tests on two influential business scenarios at Douyin\footnote{An influential short-video platform with billion user-scale: https://www.douyin.com/} are conducted to validate its performance. Currently, LONGER has been extensively developed in dozens of scenarios at ByteDance, affecting billions of users.
\end{itemize}

\section{Related Work}
\subsection{Traditional Short-Sequence Modeling}
To date, industrial recommendation systems predominantly adhere to the combined modeling paradigm of both sequence modeling and feature interaction\cite{zeng2024interformer,wang2017deep}. Within the framework, sequence modeling has long played a pivotal role in depicting user preferences. Among the extensive research, a pivotal milestone emerged with DIN \cite{zhou2018deep}. The subsequent approaches including DIEN\cite{zhou2019deep}, CAN\cite{zhou2020can}, etc. Besides, multi-domain\cite{chai2025adaptive,chang2023pepnet}, multi-interest\cite{li2019multi,chai2022user}, and sequence denoising methods\cite{shin2024attentive,chen2022denoising} are extensively approached for different aspects in modeling user preferences. Noted that most of such sophisticatedly designed structures are developed for short sequence modeling, while the long sequence modeling methods have later attracted increasing research attention.

\subsection{Long-Sequence Modeling}
As has been discussed in the Introduction, the long-sequence modeling methods can be generally categorized into two-stage retrieval, pre-trained user embedding, and memory-augmented models. Overall, the retrieval-based and pre-trained methods belong to a two-stage strategy, and the memory-enhanced models generally require long-term training periods to accumulate hit rates within the memory slots. Recently, some efforts have been made to directly model long sequence \cite{zhai2024actions, zivic2024scaling, zhang2024scaling}. However, a GPU-efficient long sequence modeling remains underexplored in large-scale industrial recommender systems.

\section{Methodology}

\begin{figure*}[htbp]  
\centering
\includegraphics[width=\textwidth,trim=50 0 50 0,clip]{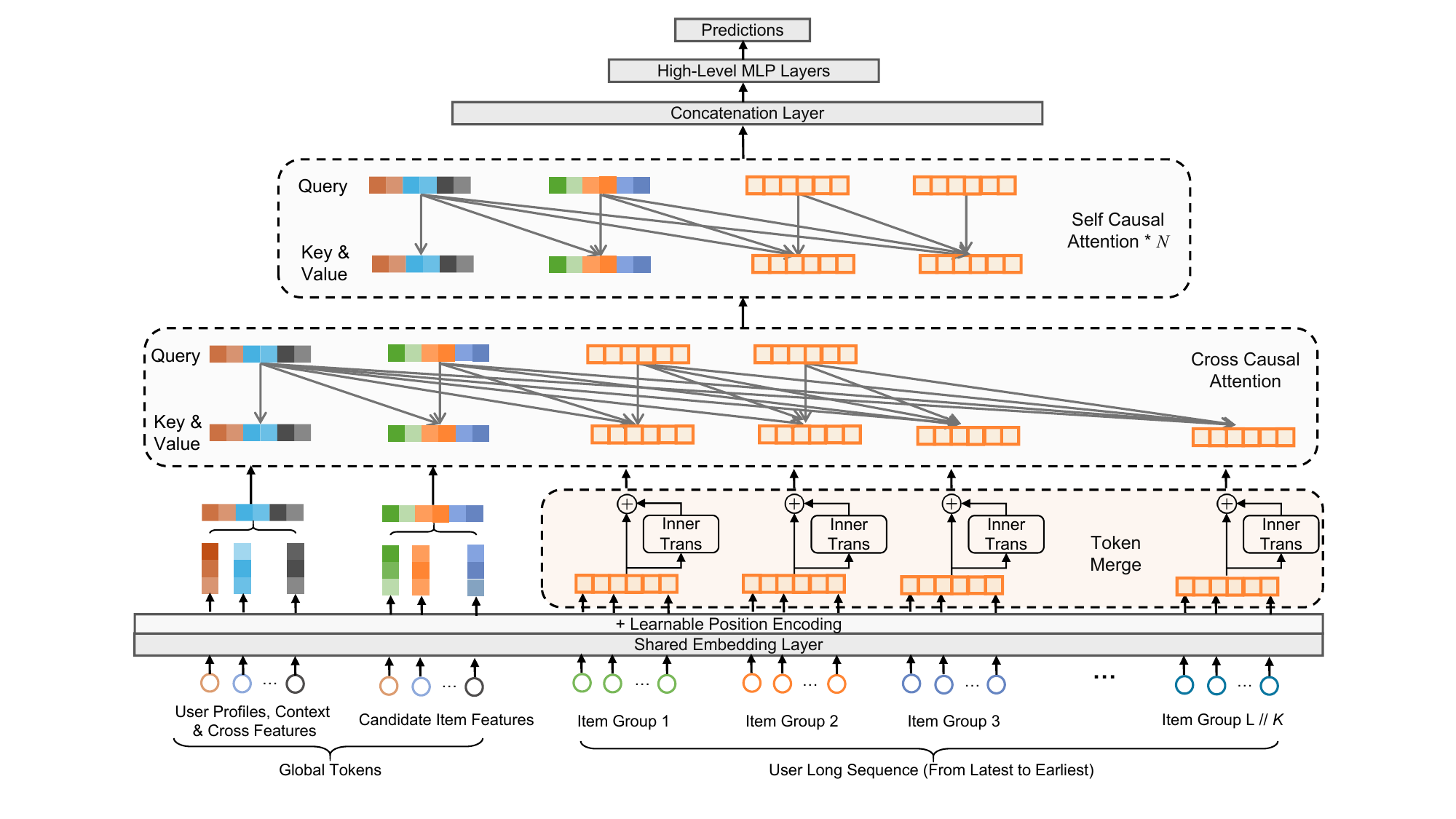}  
\caption{LONGER Model Architecture.}
\label{fig:model_image}
\end{figure*}

\subsection{Problem Statement}
Let $\mathcal{U}$ and $\mathcal{I}$ denote the user and item sets, respectively. Given a user $u \in \mathcal{U}$ with raw behavior sequence $S_u = [i^{(u)}_1, ..., i^{(u)}_L]$ where $i^{(u)}_t \in \mathcal{I}$, user basic features $u_d$ including user profiles, context features, and cross features, and a target item $v \in \mathcal{I}$, the recommendation task aims to predict the click or convert probability:

\begin{equation}
    P(y=1 \mid S_u, u_d, v) \in [0,1]
\end{equation}

\noindent where $y \in \{0,1\}$ indicates whether $u$ will interact with $v$. The model learns this mapping through historical interaction data $\mathcal{D} = \{(S_u, u_d, v, y)\}$ by optimizing the binary cross-entropy loss:

\begin{equation}
    \mathcal{L} = -\frac{1}{|\mathcal{D}|} \sum_{(S_u, u_d, v, y) \in \mathcal{D}} \left[ y \log \hat{y} + (1-y) \log (1-\hat{y}) \right]
\end{equation}

\noindent with $\hat{y} = f_\theta(S_u, v)$ being the predicted probability by the recommendation model.


\subsection{Overall Framework}
Our proposed framework is designed to address the challenges of modeling long and complex user behavior sequences in recommendation systems, while maintaining training and inference efficiency at industrial scale.  Figure~\ref{fig:model_image} illustrates the overall architecture of our proposed model, \textit{LONGER}. The framework integrates input generation, token merge, hybrid attention mechanisms, and training-serving optimizations to enable efficient and scalable long-sequence modeling.

First, we enhance the model input structure by introducing \textit{Global Tokens}, which act as aggregated anchor representations (e.g., target item representation, user ID (UID) embedding) to facilitate global information fusion and stabilize attention distributions. Next, we apply \textit{Token Merge} to compress long behavior sequences, reducing computational complexity while retaining essential local patterns. To further preserve intra-group dependencies, we introduce \textit{InnerTrans}, a lightweight inner transformer applied within merged token segments. The core model architecture, described in the \textit{LONGER Model Structure}, adopts a hybrid attention design that combines cross causal-attention (to highlight salient parts of the sequence) and stacked self causal-attention layers (to capture higher-order dependencies across the sequence).

To ensure scalability and deployment feasibility, we incorporate several engineering system-level optimizations. The framework provides fully synchronous training and serving with unified dense and sparse parameter storage on ultra-large-scale GPU clusters. We further improve memory and compute efficiency with \textit{Mixed Precision Training and Recompute}, enabling activation memory reduction and customized numerical precision. Finally, during inference, we deploy a \textit{KV Cache Serving} strategy that caches user sequence representations and reuses them across candidate scoring, significantly reducing redundant computation.

Together, these components form a cohesive system that supports long-sequence modeling with high expressiveness and efficiency, and can be readily deployed in large-scale real-world recommendation scenarios.

\subsection{Global Tokens}

We introduce \textit{Global Tokens} as auxiliary representations appended to the input sequence to facilitate global information extraction and anchoring. These tokens can include target item representation tokens, learnable CLS tokens, UID embeddings, and high-order compressed user–item interaction features. By design, global tokens possess a full attention receptive field, allowing them to aggregate contextual signals from the entire sequence while also influencing all other sequence tokens.

This architectural augmentation serves two primary purposes. First, global tokens function as centralized information anchors, enabling enhanced feature interactions between user history, contextual attributes, and candidate items. Second, they stabilize attention dynamics in long sequences, particularly under sparse attention configurations. As demonstrated in StreamLLM~\cite{xiao2024efficientstreaminglanguagemodels}, incorporating a small number of global tokens alleviates the “attention sink” effect, where deeper attention layers disproportionately focus on early tokens. These tokens act as anchor points that maintain attention diversity and preserve long-range dependency modeling.

\subsection{Token Merge}
    Let $L$ be the sequence length and $d$ be the embedding dimension. Processing long behavior sequences (typically $L \geq 2000$) with vanilla Transformers imposes prohibitive computational costs due to the quadratic attention complexity $O(L^2 d)$, especially when $L \gg d$ (typically, \ $L = 2000$, $d = 32$ in industrial recommenders). Conventional solutions such as sequence truncation lead to the loss of long-range dependencies. To address this, we propose a \textit{Token Merge} strategy that groups adjacent tokens and compresses them into shorter sequences, achieving a trade-off between model efficiency and representational fidelity. This strategy reduces the sequence length by a factor of $K$, effectively performing spatial compression. The grouped token representations can be formed via simple concatenation or further enhanced by incorporating intra-group interactions through lightweight InnerTrans blocks. This design provides a flexible trade-off between efficiency and expressiveness, preserving local semantics while enabling global modeling over a shorter sequence.
    
    Given a standard-structured transformer encoder layer, the FLOPs and parameters can be expressed as \cite{narayanan2021efficient}:

    \begin{align}
        &\text{FLOPs}_{\text{vanilla trans}} = 24Ld^2 + 4L^2d \label{eq:flops} \\
        &\text{Params}_{\text{vanilla trans}} = 12d^2 + 13d \label{eq:params}
    \end{align}
    
    \noindent\textbf{Computational Complexity.} The attention complexity ratio before and after token merge is:
    
    \begin{align}
    \frac{\text{FLOPs}_{\text{Merge Token}}}{\text{FLOPs}_{\text{vanilla}}} = \frac{24Ld^2K + \frac{4L^2d}{K}}{24Ld^2 + 4L^2d} = \frac{6dK + \frac{L}{K}}{6d + L}\notag
    \label{eq:flops_approx}
    \end{align}

    \noindent For typical $L=2048$, $d=32$:
    \begin{itemize}
        \item Vanilla Transformer: $\text{FLOPs} \approx 587 \text{M}$
        \item Merging ($K=4$): $\text{FLOPs} \approx  336 \text{M}$ (42.8\% reduction)
    \end{itemize}
    
    \noindent\textbf{Parameter Expansion} Token merging reduces computational complexity by shortening the sequence length, and simultaneously increasing the number of parameters $\Theta_{\text{merge}}$, thereby improving both efficiency and the model's expressiveness, benefiting the overall model performance.
    \begin{equation}
        \Theta_{\text{merge}} = 12K^2d^2 + 13Kd
    \end{equation}
    
    \noindent\textbf{InnerTrans.} To merge multiple adjacent tokens into one, simple concatenation of tokens within a group may result in insufficient interaction between tokens, potentially leading to the loss of fine-grained details. To address this, we introduce InnerTrans, which applies a transformer within each token group to enable local interactions. This approach ensures that the interactions within each group are effectively captured without the loss of information that typically occurs with direct concatenation. Due to the very small dimension and sequence length, the computation budget of InnerTrans is quite limited in practice.
    
    \begin{equation}
    \mathbf{M}_i = \text{TransformerBlock}\left([\mathbf{e}_{i}^1,...,\mathbf{e}_{i}^K]\right)
    \end{equation}

    \noindent where $\mathbf{M}_i$ denotes the representation of the $i$th group and $\mathbf{e}_{i}^k$ denotes the $k$th item embedding in the $i$th group.

\subsection{LONGER Model Structure}
    In our model architecture, we use a hybrid attention mechanism that combines both cross-attention and self-attention layers to efficiently process the input sequences.
    
    \subsubsection{Input Generation}
    The input to the model consists of two main components: global tokens and sequence tokens. The global tokens, which represent contextual information (such as target item features and user identifiers, as discussed in Section 3.3), are concatenated with the sequence tokens to form the input.  

    To better capture temporal dynamics in user behavior sequences, we augment the sequence tokens with additional positional side information. Specifically, two forms of positional encoding are incorporated: (1) an absolute time-difference feature that quantifies the temporal distance between each user interaction and the target item is used as side information and concatenated to each item embedding; and (2) a learnable absolute positional embedding that encodes the position of each token within the sequence which is added to the item embedding.
    
    After the position encoding, the resultant tokens are passed through a multi-layer perceptron (MLP) to generate their input representations \( \mathbf{R} \in \mathbb{R}^{(m+L) \times d} = [\mathbf{G} \in \mathbb{R}^{m \times d}; \mathbf{H} \in \mathbb{R}^{L \times d}]\) where $\mathbf{G}$ and $\mathbf{H}$ denote the global token and sequence token representations, respectively. The query matrix \( \mathbf{O} \) is then constructed by concatenating $m$ global tokens \( \mathbf{G} \in \mathbb{R}^{m \times d} \) with the $k$ sampled sequence tokens \( \mathbf{H}_\mathbf{S} \in \mathbb{R}^{k \times d} \), which are selected from the full sequence tokens $\mathbf{H} $ based on a predefined sampling strategy. Similar query compression ideas have also been explored in other research fields, for example, Perceiver\cite{jaegle2021perceiver} and Q-Former\cite{ li2023blip}, which adopt a learnable token strategy for compression. In experiments, we comprehensively compare different strategies, including taking the most \textit{recent \( k \)} or \textit{uniformly sampled tokens}, or initialize \( k \) \textit{learnable tokens}, and find that \textit{recent \( k \)} provides the best results. This hybrid attention design is also motivated by the observation that model performance exhibits strong marginal effects with respect to the number of sequence tokens: sampling just 40\% of the full sequence retains over 95\% of the performance improvement, while reducing around 50\% FLOPs (see Section 4). The composite query is then constructed as:
    
\begin{equation}
\mathbf{O} = [\mathbf{G}; \mathbf{H}_\mathbf{S}]
\end{equation}
    
    \noindent This hybrid design focuses attention on both critical local behaviors and global contextual signals, enabling the model to efficiently capture both specific sequence dependencies and broader contextual information.
    
    \subsubsection{Cross-Causal Attention (First Layer)} 
    In the first attention layer, we apply cross-causal attention using the query matrix \( \mathbf{O} \) generated in the previous step, and the input tokens \( \mathbf{R} \in \mathbb{R}^{(m+L) \times d} \). The cross-attention mechanism is computed as:
    
    \begin{equation}
    \mathbf{Q} = \mathbf{O} \mathbf{W}_\mathbf{Q}, \quad \mathbf{K} = \mathbf{R} \mathbf{W}_\mathbf{K}, \quad \mathbf{V} = \mathbf{R} \mathbf{W}_\mathbf{V} 
    \end{equation}

    \begin{equation}
    \text{Attention}(\mathbf{Q}, \mathbf{K}, \mathbf{V}) = 
    \text{Softmax}\left(\frac{\mathbf{Q} \mathbf{K}^T}{\sqrt{d}} + \mathbf{M}\right)\mathbf{V}
    \end{equation}
    
    \noindent where $\mathbf{W}_\mathbf{Q}$, $\mathbf{W}_\mathbf{K}$ and $\mathbf{W}_\mathbf{V}$ represent the query, key, and value projections with shape $\mathbb{R}^{d \times d}$, and the mask matrix $\mathbf{M}$ is defined as:
    
    \begin{equation}
    \mathbf{M}_{i,j} =
    \begin{cases}
    0, & \text{if }  j \geq i,\ \text{where} \{i, j\} \in [1, m+L] \\
    -\infty, & \text{otherwise}
    \end{cases}
    \end{equation}

    The causal mask design, on one hand, maintains temporal relevance between sequence items. On the other hand, it ensures the invisibility from the sequence to the candidate item, enabling the KV Cache Serving mechanism (see Section 3.6.3). After computing the attention, the result is passed through a feed-forward network (FFN) for further processing. 
    
    \subsubsection{Self-Causal Attention (Subsequent Layers)} 
    After the cross-causal attention layer, the subsequent layers consist of several self-causal attention blocks. These layers focus on learning the internal relationships within the sampled tokens sequence, allowing the model to capture dependencies and patterns within the tokens of the behavior sequence itself. Each self-causal attention layer is followed by an FFN, which helps in further processing the information learned by the attention mechanism. The self-causal attention mechanism is computed using a similar formulation:
    
    \begin{equation}
            \text{SelfAttention}(\mathbf{Q}, \mathbf{K}, \mathbf{V}) = \text{softmax}\left(\frac{\mathbf{Q}\mathbf{K}^T}{\sqrt{d}}   +  \mathbf{M} \right)\mathbf{V}
    \end{equation}
    
    \noindent Here, the query, key, and value are obtained by applying separate linear projections $\mathbf{W}_\mathbf{Q}$, $\mathbf{W}_\mathbf{K}$, $\mathbf{W}_\mathbf{V}$ to the output of the previous layer.

    \subsubsection{Stacking and Compression} 
    The self-causal attention layers are stacked \(N\) times to iteratively refine the representations of the input sequence. After passing through these layers, the model produces a compressed output, which represents the final output of the attention mechanism. This output is then used for the downstream prediction task.
    
    \begin{equation}
        \underbrace{\text{CrossAttn}(\mathbf{O}, \mathbf{R})}_{\mathclap{\text{compress long sequence}}} \longrightarrow \underbrace{\text{SelfAttn}(\cdot)\times N}_{\mathclap{\text{high-order interactions}}}
        \end{equation}
    
    By using a combination of cross-attention in the first layer and self-attention in subsequent layers, our model is able to efficiently handle long sequences while leveraging both global context and internal dependencies.

\subsection{Training and Deployment Optimization}
\subsubsection{Training Framework} 

\begin{figure}[H]
    \centering
    \includegraphics[width=1\linewidth,trim=0 70 0 75,clip]{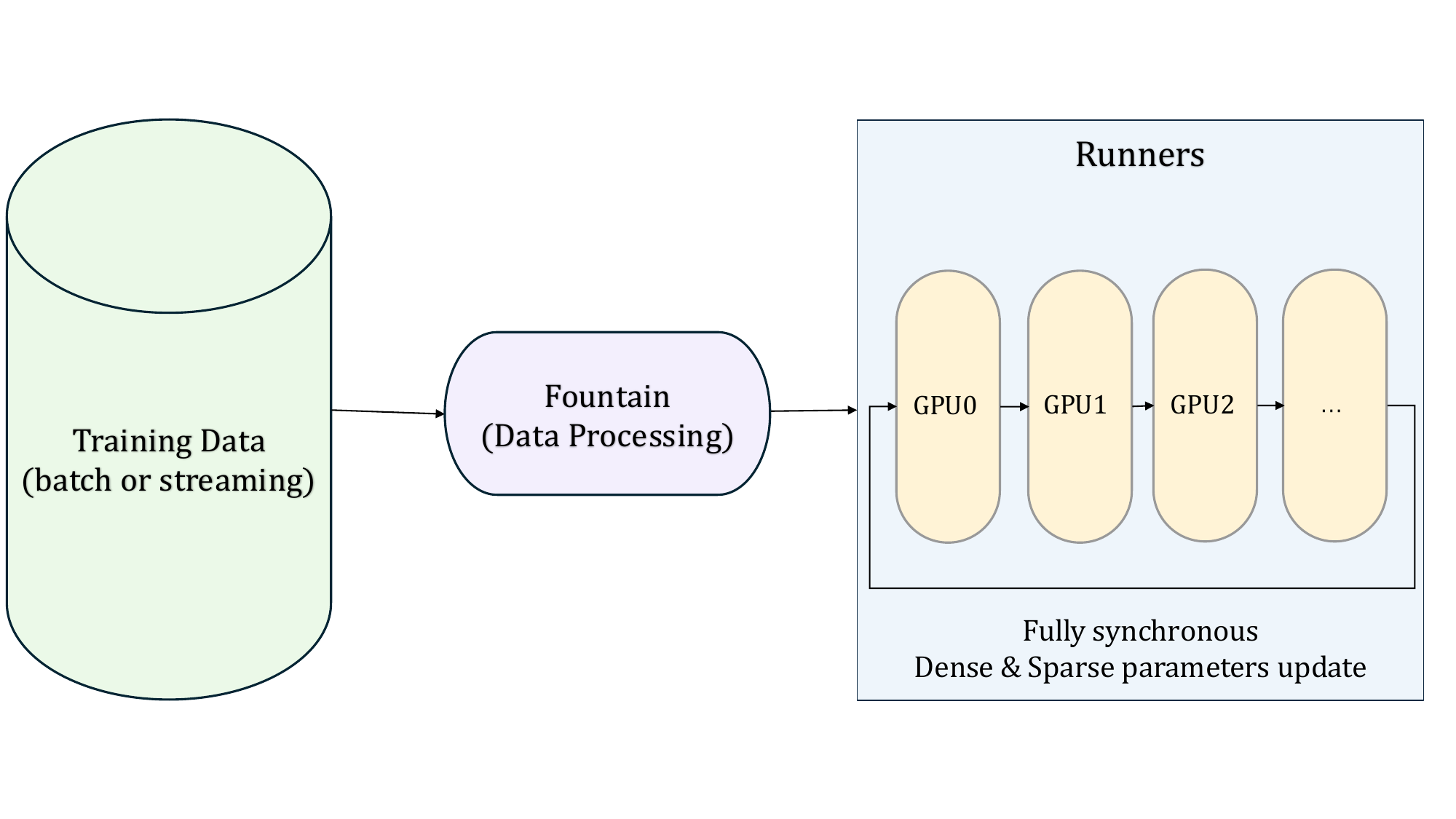}
    \caption{Training  Framework}
    \label{fig:jaguar}
\end{figure}
Our training framework is a fully synchronous system designed for large-scale sparse models, tailored to leverage the capabilities of modern high-performance GPUs. Built upon a hardware–software co-design philosophy, it aims to maximize computational throughput and memory efficiency in distributed training. The training pipeline begins with data ingestion in batch or streaming form, followed by preprocessing through the \textit{Fountain} module. The processed training data are then dispatched to multiple GPU runners, where both dense and sparse parameters are updated synchronously. This unified design facilitates effective scaling across devices and nodes, providing a robust foundation for training large-parameter models in production environments.

A defining characteristic of the framework is its unified parameter storage and training architecture. Both dense and sparse parameters are stored and updated synchronously on GPU machines, eliminating the need for external Parameter Server components. To better accommodate the feature distribution patterns in recommendation systems, the framework adopts a hierarchical memory system for sparse embedding, enabling efficient support for large embedding tables. In this design, high-frequency features are stored in high-bandwidth GPU memory (HBM), mid-frequency features reside in CPU main memory (MEM), and low-frequency features are offloaded to local solid-state drives (SSD). This stratified storage layout is optimized to match the access characteristics of recommendation data, providing a practical trade-off between latency, throughput, and capacity. The core innovation lies in fully colocating both computation and parameter storage on the GPU machines, thereby reducing communication overhead and memory transfer latency. This results in improved training throughput, reduced staleness, and enhanced convergence stability.

\subsubsection{Mixed Precision Training and Recompute}

To alleviate GPU memory pressure during training, we adopt a recompute strategy alongside mixed precision training. For gradient computation, we use reverse-mode automatic differentiation, which is more efficient than forward-mode but requires storing all intermediate activations from the forward pass. These activations can become a major memory bottleneck. To address this, we support recomputing declarations at the model definition level, allowing selected activations to be discarded during the forward pass and recomputed during the backward pass. This trades computation for memory savings. As native TensorFlow does not provide official support for recomputation, we implement it using the \texttt{custom\_gradient} mechanism, enabling fine-grained control through code-level annotations.

In addition, to reduce compute overhead caused by dense model scaling, we adopt BF16/FP16-based mixed precision training. Users can configure precision at the model level, applying higher precision to critical components and lower precision elsewhere. This approach has shown substantial benefits in production workloads, including +18\% throughput, -16\% training time, and -18\% memory usage on average, with up to -28\% memory reduction in dense layers.

\subsubsection{KV Cache Serving}
To improve inference efficiency when scoring multiple candidates, motivated by M-FALCON \cite{zhai2024actions}, we introduce a KV caching mechanism that decouples the attention computation between user behavior tokens and the candidate-specific global token. Since the user sequence remains the same across candidates, its internal representation can be computed once and reused.

Specifically, we split the attention input into two parts: (1) the user sequence tokens, and (2) the global token associated with the candidate item. The key and value projections of the user sequence are precomputed and cached. For each candidate, only the attention involving its global token and the cached user sequence is computed. This leads to a two-stage inference process:
\begin{enumerate}
  \item Precompute and cache the key-value tensors of the user sequence.
  \item Compute attention between each candidate's global token and the cached user sequence.
\end{enumerate}

\begin{figure}[tbp]
    \centering
    \includegraphics[width=1\linewidth,trim=0 70 0 75,clip]{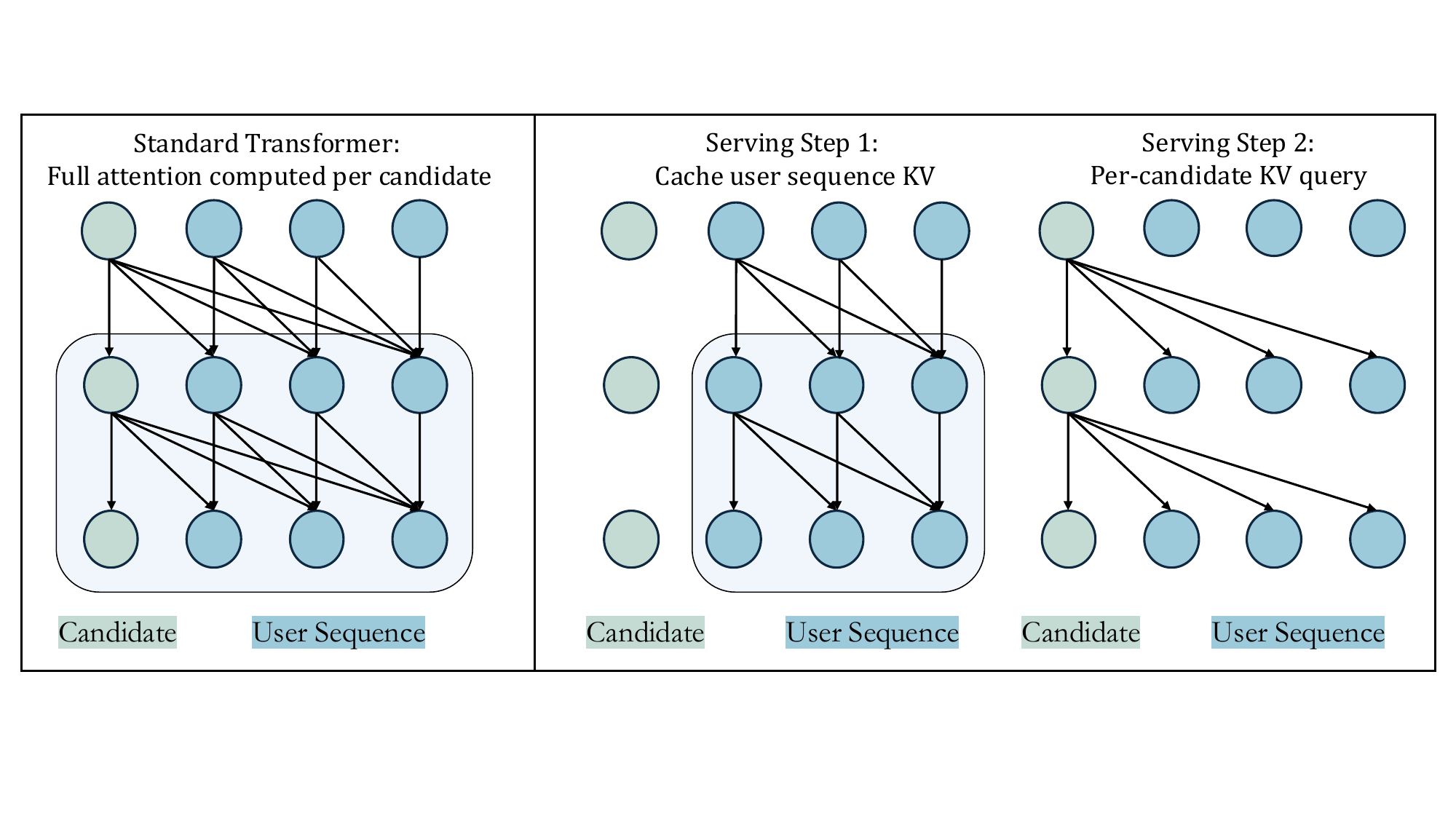}
    \caption{KV Cache Serving}
    \label{fig:kvserving}
\end{figure}

As shown in Figure~\ref{fig:kvserving}, this optimization avoids redundant computation and significantly reduces serving latency. In practice, it improves online serving efficiency, reducing throughput degradation from as high as \(-40\%\) to only \(-6.8\%\).

\begin{table*}[ht]
    \centering
    \caption{Evaluation of methods on industrial datasets}
    \label{tab:performance}
    \begin{tabular}{l l l l l l l l l}
        \toprule
          & Base& SumPooling  & TWIN & DIN (Recent50) & DIN  & HSTU & Transformer & \textbf{LONGER} \\
        \midrule
        \textbf{AUC$\uparrow$} & 0.83968 & 0.84201 & 0.84472  & 0.84698 & 0.84982 & 0.84994& 0.85111& \textbf{0.85290}\\
        \textbf{LogLoss$\downarrow$} & 0.48758 & 0.48538 & 0.48168  & 0.47830 & 0.47452 & 0.47490 & 0.47293 & \textbf{0.47103}\\
        \textbf{$\Delta$AUC(\%)} & - & +0.28 & +0.60  & +0.87 &+1.21 & +1.22& +1.36 & \textbf{+1.57}\\
        \textbf{$\Delta$LogLoss(\%)} & - & -0.45  & -1.21  & -1.90 & -2.68 & -2.60 & -3.00 & \textbf{-3.39}\\
        \bottomrule
    \end{tabular}
\end{table*}

\section{Experiments}

\subsection{Experimental Setting}
We evaluate our model on the Conversion Rate (CVR) prediction task in the Douyin Ads system, a real-world, large-scale industrial advertising recommendation scenario. The dataset is constructed from a subset of online user interaction logs collected between October 16th, 2024 and February 23rd, 2025, comprising 5.2 billion samples over 130 consecutive days. Each sample includes user demographic features like user ID (UID), gender, ultra-long user behavior sequence, and a candidate ad item. The user behavior sequences contain various interaction types, including page views, clicks, and conversions, while item-side features cover ad content, display context, and associated metadata. We adopt a temporally consistent data split strategy: the first 123 days are used for training, and the remaining 7 days are reserved for offline evaluation. This setup aligns with real-world deployment practices and effectively prevents future data leakage during model development.

For comparison, we evaluate our model against several strong baselines, categorized by their ability to model short- or long-range user behavior. Short-sequence methods include TWIN \cite{chang2023twin} and DIN (Recent50) which rely on 50 interactions. Long-sequence methods, including SumPooling, DIN\cite{zhou2018deep}, HSTU \cite{zhai2024actions}, and Transformer\cite{chen2019behavior}, process extended behavior histories that often suffer from scalability and efficiency issues in industrial environments. All models are trained with the same preprocessing pipeline and hyperparameter tuning, and experiments are conducted on a 48×A100s GPU cluster.

\begin{table}[ht]
\footnotesize	
\setlength{\tabcolsep}{5pt}  
\renewcommand{\arraystretch}{1.25}
\caption{Ablation Study on Query Quantity and Key Components of LONGER.}
\label{tab:ablation_combined}
\centering
\scalebox{0.92}{
\begin{tabular}{lccccc}
\toprule
\textbf{Configuration} & \textbf{\makecell{FLOPs\\($\times10^9$)}} & \textbf{AUC$\uparrow$} & \textbf{LogLoss$\downarrow$} & \textbf{$\Delta$AUC} & \textbf{$\Delta$LogLoss} \\
\midrule
LONGER  (w/o Merge, 2000)           & 3.73 & 0.85111 & 0.47293 & +1.36\% & -3.00\% \\
+TokenMerge4(Concat, 500)          & 2.13 & 0.85232 & 0.47145 & +1.51\% & -3.31\% \\
+TokenMerge8(Concat, 250)          & 3.03 & 0.85291 & 0.47062 & +1.58\% & -3.48\% \\
\midrule
\multicolumn{5}{l}{\textit{Based on LONGER with TokenMerge8}} \\
\quad + InnerTrans           & 3.52 & 0.85332 & 0.47052 & +1.63\% & -3.50\% \\
\midrule
\multicolumn{5}{l}{\textit{Varying Query Number (Sampling Recent $k$ items)}} \\
\quad Query number = 50                   & 1.27 & 0.85235 & 0.47162 & +1.51\% & -3.27\% \\
\quad Query number = 80                   & 1.59 & 0.85248 & 0.47157 & +1.52\% & -3.28\% \\
\quad \textbf{Query number = 100}         & \textbf{1.91} & \textbf{0.85290} & \textbf{0.47103} & \textbf{+1.57\%} & \textbf{-3.39\%} \\
\quad Query number = 150                  & 2.36 & 0.85290 & 0.47101 & +1.57\% & -3.40\% \\
\quad Query number = 200                  & 2.93 & 0.85331 & 0.47077 & +1.62\% & -3.45\% \\
\quad Query number = 250                  & 3.52 & 0.85332 & 0.47052 & +1.63\% & -3.50\% \\
\midrule
\multicolumn{5}{l}{\textit{Query Selection Strategies}} \\
\quad  Learnable 100 & 1.91 & 0.84946 & 0.47523 & +1.17\% & -2.53\% \\
\quad  Recent 100     &1.91 & 0.85290 & 0.47103 & +1.57\% & -3.39\% \\
\quad  Uniform 100   & 1.91 & 0.85183 & 0.47215 & +1.45\% & -3.16\% \\
\quad  Recent50 + Rest Unif50  & 1.91 & 0.85255 & 0.47129 & +1.53\% & -3.34\% \\

\bottomrule
\end{tabular}}
\end{table}

\subsection{Overall Performance}

\subsubsection{Comparison of existing methods.} We report model performance on the offline evaluation set using two standard metrics for binary classification in recommendation systems: AUC (Area Under the ROC Curve) and LogLoss. Table ~\ref{tab:performance} summarizes the results across multiple baselines and our proposed model. According to the table, our model outperforms all baselines, achieving an AUC of 0.85290 and a LogLoss of 0.47103, which represents a relative improvement of 1.57\% in AUC  compared to the base model, and improves the AUC by 0.21\% compared to the most competitive model, i.e., Transformer. It is noted that a 0.1\% improvement is considered to be a significant improvement that can affect the performance in online A/B test in the industrial case. Besides, the proposed model also demonstrates significantly higher efficiency compared to vanilla Transformer (see Section 4.2.2). This improvement demonstrates the effectiveness of our approach in capturing long-range behavior dependencies while maintaining computational efficiency.

\subsubsection{Ablation study.} Table~\ref{tab:ablation_combined} presents an ablation study on the key components and query-related configurations in \textit{LONGER}. We first examine the impact of the TokenMerge module and the InnerTrans component. Compared to the base model without merging, integrating TokenMerge (Concat, 250) reduces FLOPs from $3.73 \times 10^9$ to $3.03 \times 10^9$, while improving AUC by 1.58\% and decreasing LogLoss by 3.48\%. Further incorporating InnerTrans brings additional gains, achieving the best overall LogLoss of 0.47052 and a 1.63\% AUC improvement.

Next, we vary the number of queries ($k$) used to summarize recent user behaviors. The results show that increasing $k$ generally improves performance but also increases computation. Notably, using 100 queries achieves a strong trade-off, with an AUC of 0.85290 and a LogLoss of 0.47103—very close to the performance obtained when using all queries ($k=250$), but with only 54\% of the FLOPs. This setting is highlighted in bold in Table~\ref{tab:ablation_combined}, showing its practicality for real-world deployment where computational budgets are critical.

Finally, we compare different query selection strategies. These strategies can be viewed as different initialization methods for the query set. Among them, using learnable queries (initialized randomly) performs the worst (AUC = 0.84946). In contrast, directly selecting the most recent 100 user behaviors (Recent 100) achieves the best overall performance. Other strategies, such as uniform sampling or combining recent and uniformly sampled items, yield slightly lower AUC and higher LogLoss. These findings suggest that initializing queries with informative behaviors—particularly recent ones—is crucial for effectively capturing user intent in long-sequence modeling.

Overall, the ablation study confirms that both architectural enhancements (e.g., TokenMerge, InnerTrans) and query-related strategies (e.g., query number and selection method) play critical roles in balancing accuracy and efficiency. The findings validate that \textit{LONGER} can achieve strong performance with reduced computational cost by carefully designing its key components and behavior modeling pipeline. Such a configuration makes our method highly suitable for large-scale industrial deployment, where low-latency inference and system throughput are essential.

\subsection{Scaling Analysis}

In this section, we present the scaling analysis of model performance with respect to sequence length, FLOPs, and the number of parameters. The scaling behavior of these factors follows the general form:

\begin{equation}
    y = \alpha x^\beta + \gamma
\end{equation}

\noindent where \( y \) represents the performance metric (AUC and LogLoss), \( x \) represents the scaling factor (sequence length, FLOPs, or parameters), \( \alpha \) and \( \beta \) are constants, and \( \gamma \) represents a constant offset.
\subsubsection{Sequence Length}
\begin{figure*}[ht]
    \centering
    \begin{minipage}{0.48\linewidth}
        \centering
        \includegraphics[width=1\linewidth ]{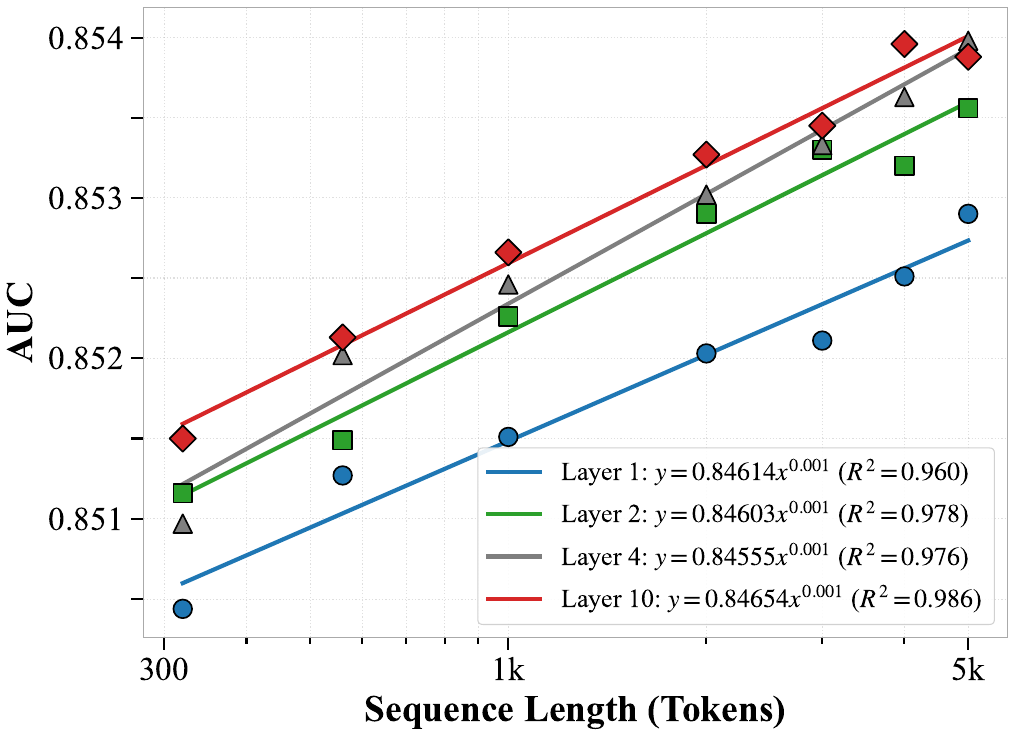}
    \end{minipage}%
    \hfill
    \begin{minipage}{0.48\linewidth}
        \centering
        \includegraphics[width=1\linewidth]{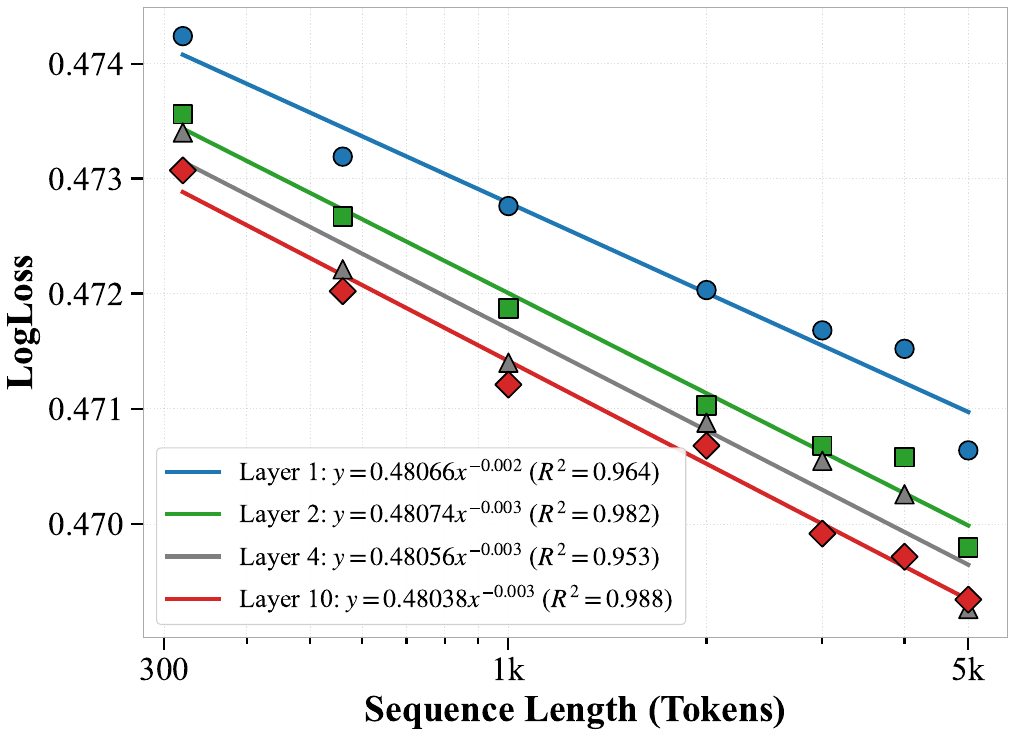}
    \end{minipage}
    \caption{Scaling up sequence length in LONGER.}
    \label{fig:length}
\end{figure*}
We analyze how performance scales with input sequence length across different model depths. As shown in Figure ~\ref{fig:length}, increasing the number of tokens consistently improves AUC and reduces LogLoss, following a power-law trend. Deeper models benefit more from longer sequences, but AUC improvement slows with depth, indicating diminishing returns. The optimal depth should balance model capacity and computational constraints.

Overall, longer sequences enhance performance, especially when paired with an appropriately chosen depth. Beyond a certain depth, further gains are marginal.

\begin{figure}[H]
  \centering
  \begin{subfigure}[b]{0.48\linewidth}
    \centering
    \includegraphics[width=\linewidth]{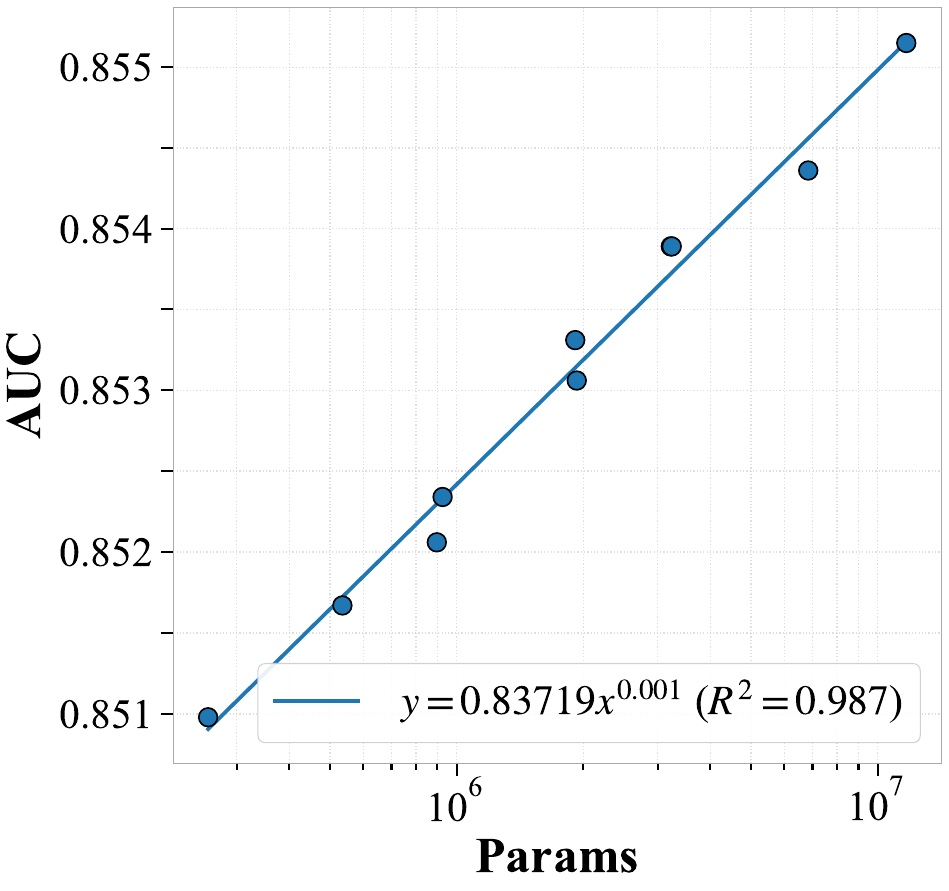}
    \caption{Params vs. AUC}
    \label{fig:flops_auc}
  \end{subfigure}
  \hfill
  \begin{subfigure}[b]{0.48\linewidth}
    \centering
    \includegraphics[width=\linewidth]{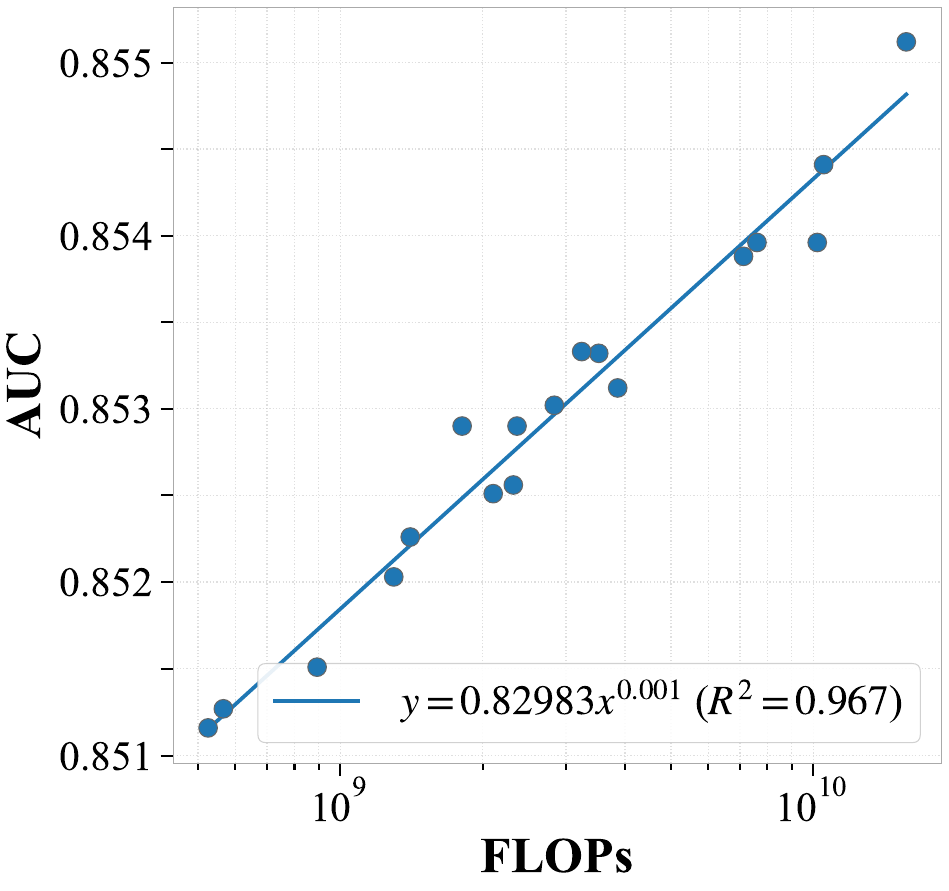}
    \caption{FLOPs vs. AUC}
    \label{fig:params_auc}
  \end{subfigure}
  \caption{Scaling performance with respect to FLOPs and model parameters.}
  \label{fig:scaling_flops_params}
\end{figure}

\subsubsection{Parameters}

We evaluate model capacity by scaling the hidden dimension size while fixing the number of layers to 2 and the input sequence length to 2000. As shown in Figure~\ref{fig:scaling_flops_params}(a), AUC increases steadily with parameter count, following a strong power-law trend ($R^2 = 0.987$). These results demonstrate that increasing model width effectively enhances performance under fixed architecture, with no sign of saturation in the current parameter range.

\subsubsection{FLOPs}

We analyze model performance by varying the number of layers and sequence length while keeping the model dimensionality fixed at 32. As shown in Figure~\ref{fig:scaling_flops_params}(b), AUC increases steadily with FLOPs, following a strong power-law trend ($R^2 = 0.967$). This indicates that increasing computational resources enables the model to process longer or more complex user behavior sequences, capturing higher-order dependencies and improving prediction accuracy, even under a fixed model width.

These results suggest that increasing computational resources is an effective way to improve performance, but the efficiency gain should be balanced against the computational and memory constraints typically encountered in real-world systems.

\subsection{Online A/B Tests}
In this section, we present the results of the online A/B tests, which were conducted to evaluate the effectiveness of the proposed model in real-world scenarios within both Douyin Ads and Douyin E-Commerce Platforms, both of which are very influential commercial platforms and attract billions of users. The baseline models in these scenarios are already quite strong, making the observed improvements even more significant. The dual-domain testing allowed us to evaluate the model's generalization ability in both advertising and e-commerce environments, which are critical components of the platform's ecosystem.

\subsubsection{Douyin Ads Platform.}
This section presents the results of the A/B test for Douyin Ads, where we evaluate the performance of our model using two key metrics: ADSS (Advertiser Score) and ADVV (Advertiser Value), which are the most important indicators in industrial advertising systems. The test was conducted across three advertisement formats: Live Streaming, Short Video, and Mall. For Live Streaming, the model achieved a 1.063\% improvement in ADSS and a 1.168\% improvement in ADVV. In the Short Video format, ADSS is increased by 2.097\%, while ADVV showed a 2.151\% improvement. In the Mall format, ADSS is improved by 1.816\%, and ADVV was increased by 1.407\% . These results confirm that the model effectively enhances performance across all advertisement formats with consistent improvements.
\begin{table}[htbp]
\centering
\caption{Douyin Ads A/B Test Results}
\label{tab:ab_test_results}
\begin{tabular}{lcc}
\toprule
\textbf{Advertise Type} & \textbf{ADSS} & \textbf{ADVV} \\
\midrule
Live Streaming & +1.063\% & +1.168\% \\
Short Video    & +2.097\% & +2.151\% \\
Mall           & +1.816\% & +1.407\% \\
\bottomrule
\end{tabular}
\end{table}

\subsubsection{Douyin E-Commerce Service.}

 For the A/B test on Douyin E-Commerce, we evaluate the effectiveness of different content formats using two key metrics: Order/U (the number of orders per user) and GMV/U (the gross merchandise volume per user). These metrics help us understand the impact of the model not only on total sales volume but also on user-level engagement and value generation. The results show significant improvements in both metrics. For Live Streaming, Order/U is increased by 7.9222\%, and GMV/U is lifted by 6.5404\%, indicating that live streaming contents have a strong positive effect on both the number of orders and the value generated per user. In the Short Video content, Order/U is improved by 4.6125\%, and GMV/U is increased by 5.2771\%, demonstrating the effectiveness of short video content in boosting overall sales per user. These results highlight the substantial impact of both ad formats, with Live Streaming showing notably larger improvements in both Order/U and GMV/U.

\begin{table}[htbp]
\centering
\caption{Douyin E-commerce A/B Test Results}
\label{tab:ab_test_results}
\begin{tabular}{lcc}
\toprule
\textbf{E-commerce Type} & \textbf{Order / U} & \textbf{GMV / U} \\
\midrule
Live Streaming & +7.9222\% & +6.5404\% \\
Short Video    & +4.6125\% & +5.2771\% \\
\bottomrule
\end{tabular}
\end{table}

\section{Conclusions}

In this paper, we presented \textbf{LONGER}, a Transformer-based framework designed for efficient and scalable modeling of ultra-long user behavior sequences in industrial recommender systems. By introducing a series of architectural designs including global tokens, token merge with InnerTrans, hybrid causal attention, and system-level optimizations including the GPU-synchronous framework, mixed-precision and recomputation training, and KV cache serving, LONGER enables end-to-end ultra-long sequence modeling under real-world industrial constraints. Extensive experiments on industrial billion-scale datasets and online A/B tests across both advertising and e-commerce domains validate its robustness and generalizability at billion-user industrial scale. Notably, LONGER achieves competitive accuracy while significantly reducing computational overhead, making it well-suited for deployment in latency-sensitive production environments. Future work includes investigating more efficient sequence modeling techniques and improving cross-domain behavior modeling in industry.




\balance
\bibliographystyle{ACM-Reference-Format}
\bibliography{ref}


\end{document}